\documentclass[runningheads]{llncs}
\usepackage{cite}
\usepackage{graphicx}
\usepackage{hyperref}
\usepackage{amsmath}
\usepackage{amssymb} 
\usepackage{algorithmic}
\usepackage{enumitem}
\usepackage{cleveref}
\usepackage{color}
\usepackage{booktabs}
\usepackage{multirow,bigdelim}
\usepackage{pgfplots}
\usepackage{mathtools}
\usepackage{tikz}
\usepackage{algorithm}
\usepackage{algorithmic}
\usepackage[binary-units=true]{siunitx}
\usetikzlibrary{shapes.geometric, arrows, calc}
\tikzstyle{state} = [rectangle, rounded corners, minimum width=5cm, minimum height=0.5cm,text centered, very thick,  draw=black, fill=red!50]
\tikzstyle{arrow} = [thick,->,>=stealth]

\pgfplotsset{compat=1.16}

\usepackage[caption=false,font=footnotesize]{subfig}

\usepackage{todonotes}
\newcommand{\self}{\ensuremath{\operatorname{self}}}

\begin{document}

\title{Shared-Dining: Broadcasting Secret Shares using Dining-Cryptographers Groups}
\titlerunning{Shared-Dining}

\author{David Mödinger\inst{1}\and
	Juri Dispan\inst{1}\and
	Franz J. Hauck\inst{1}}
\authorrunning{D. Mödinger et al.}
%
\institute{Institute of Distributed Systems, Ulm University, 89081 Ulm, Germany\\
	\email{\{david.moedinger, juri.dispan, franz.hauck\}@uni-ulm.de}
}

\maketitle

\begin{abstract}
A k-anonymous broadcast can be implemented using a small group of dining cryptographers to first share the message, followed by a flooding phase started by group members. 
Members have little incentive to forward the message in a timely manner, as forwarding incurs costs, or they may even profit from keeping the message.
In worst case, this leaves the true originator as the only sender, rendering the dining-cryptographers phase useless and compromising their privacy.
We present a novel approach using a modified dining-cryptographers protocol to distributed shares of an $(n,k)$-Shamir's secret sharing scheme.
Finally, all group members broadcast their received share through the network, allowing any recipient of $k$ shares to reconstruct the message, enforcing anonymity.
If less than k group members broadcast their shares, the message cannot be decoded thus preventing privacy breaches for the originator.
%
Our system provides \((n-|\text{attackers}|)\)-anonymity for up to $k-1$ attackers and has little performance impact on dissemination.
We show these results in a security analysis and performance evaluation based on a proof-of-concept prototype.
Throughput rates between 10 and 100\,kB/s are enough for many real applications with high privacy requirements, e.g., financial blockchain system.
\end{abstract}

\section{Introduction}

In recent years, the general public has become more interested in privacy issues,
even leading to strong privacy-protection regulation, e.g., the general data protection regulation (GDPR) of the European Union.
This increased interest led to a rekindling of privacy research, especially for financially-sensitive information.

Several cryptocurrencies attempt to provide unlinkable transactions for their users~\cite{zcoin, monero}.
Unfortunately, many of these approaches neglected the underlying network's privacy and focused on the public information accessible through the blockchain.
Researchers showed that transactions can still be deanonymized through the network~\cite{Koshy, Biryukov}.
This network deanonymization led to even better identification, as internet-protocol (IP) addresses can be matched to real-world identities compared to public keys.

Various projects tackled this issue of network identification.
Monero~\cite{monero} applies Kovri\footnote{See \url{https://gitlab.com/kovri-project/kovri}.}, a garlic-based routing scheme.
In previous work, we proposed a protocol based on dining-cryptographers (DC) groups to realize a broadcast protocol with strong privacy guarantees~\cite{flex}.
Chaum's dining-cryptographers groups~\cite{dcns} have been used by other state-of-the-art protocols such as Dissent~\cite{dissent,dissent2} and k-anonymous groups~\cite{ahn}.

Although DC groups provide very strong privacy, their efficient usage for broadcast communication requires additional protocols layered on top of the DC network, e.g., a flood-and-prune broadcast.
This creates additional risks, as non-cooperating participants in the layered protocol might force the true originator to step up and jeopardize their anonymity.
In previous systems, timeouts were used to detect nodes responsible to broadcast but failed to do so.
Groups then had to punish or exclude these misbehaving nodes.
A better system would incentivize nodes to participate instead of only punish when misbehaving.
Proper incentives become even more important under stricter scrutiny, as misbehaving nodes might refuse cooperation selectively or drag out processes unnecessarily, leaving the true originator to forfeit their anonymity guarantees and start the flooding themselves.
Therefore, we designed a system where messages can only be read when enough participants cooperate to cross a threshold, enforcing the anonymity guarantees of the protocol throughout the network.

%
Our contribution is a novel system combining dining-cryptographers groups and $(n,k)$-Shamir's secret sharing. 
Our system prevents identification of the originator in the presence of up to $k-1$ attackers in the DC group for a given security parameter $k < n$ with a DC group size of $n$.
Broadcasting the shares requires at least k participants, leading to enforced k-anonymity during the broadcast.
Lastly, we provide a proof-of-concept implementation and its evaluation.

%
The structure of this paper is as follows:
In \Cref{sec:bkg}, we give an overview of the basic building blocks and the background of this paper.
We propose our k-resistant solution to broadcast messages using a DC-protocol and Shamir's secret sharing in \Cref{sec:scheme}.
We provide proof of our scheme's security and privacy in \Cref{sec:priv}, while an evaluation of the performance of our scheme can be found in \Cref{sec:perf}.
Lastly, in \Cref{sec:app}, we discuss possible applications of our scheme.

\section{Background}
\label{sec:bkg}

In this section, we discuss the required background for this paper.
First and foremost, this encompasses the notation, scenario, and attacker model and the algorithmic and mathematical concepts used in this paper, i.e., Chaum's dining-cryptographers protocol and Shamir's secret-sharing scheme.

\subsection{Notation and Scenario}

For this paper, we will restrict the discussion to groups of nodes that interact as peers, e.g., a peer-to-peer network.
Hereby, the network is further segregated into a group of $n$ participants, who form a group $G = g_1,\ldots,g_n$.
Each participant $g_i$ is identified by its index $i$.

Participants create various messages.
The message a participant $g_i$ creates and wants to broadcast is denoted by $m_i.$
Intermittent messages created to be sent throughout the protocol by $g_i$ and received by $g_k$ are denoted as $M_i$.
Throughout the paper, we use $\oplus$ to denote the bitwise XOR.

The group has various requirements for their network communication.
A group needs pairwise authenticated connections between all nodes to prevent network manipulation.
Further, nodes need to be able to create a securely shared secret between each pair of nodes.
The assumptions are easily satisfied by modern networks using mTLS and generally available cryptographic libraries.

\subsection{Dining-cryptographers Protocol}

Chaum's dining-cryptographers protocol~\cite{dcns} allows a participant in a group to broadcast a message with perfect sender anonymity.
This means that an attacker attempting to identify the sender of a message deducts that all non-colluding participants have an equal probability of being the sender of the message.

Conceptually, the dining-cryptographers protocol performs a distributed computation of the bitwise XOR function $\bigoplus_{i=1\dots n} m_i$ where each participant provides one input value $m_i$.
In case participant $g_k$ is sending a message $m_k$ and every other participant is using $m_{i\not=k}=0$, each member computes 
\begin{equation}
m_{out} = \bigoplus_{i\in 1\ldots n} m_i = 0 \oplus 0 \oplus \dots \oplus m_{k} \oplus \dots \oplus 0 = m_{k}.
\end{equation}

To compute a bitwise XOR, and therefore hide the true sender, all messages need to have the same length.
This requirement can be lifted by application of preparing communication steps as used by Dissent~\cite{dissent2}.
At most, one message is allowed to be non-zero, otherwise, the resulting message $m_{out}$ would be the XOR of all input messages and therefore unreadable.
A node that does not intend to send anything uses $m_i=0$ as an input message.
The protocol as described in \Cref{alg:dcn1} is run by every node separately, broadcasting one message per-protocol run.

\begin{algorithm}[htbp]
\begin{algorithmic}[1]
	\renewcommand{\algorithmicrequire}{\textbf{Input:}}
	\renewcommand{\algorithmicensure}{\textbf{Output:}}
	\REQUIRE Participants $g_1, g_2, \dots, g_n$, message $m_{\self}$ of length $\ell$
	\ENSURE  Message $m_{out}=\bigoplus_{k=1\ldots n} m_k$ which is the same across all participants
	\STATE\label{phone} Establish shared random secrets $s_{\self,i}$ of length $\ell$ with each member $g_i,i\not=\self$
	\STATE $M_{\self} = m_{\self} \oplus \bigoplus_{i=1\dots n,i\not=\self} s_{\self,i} $
	\STATE\label{step4} Send $M_{\self}$ to $g_i$ $\forall i \in \{1\dots n\}\setminus\{\self\}$
	\STATE\label{s5a} Receive $M_i$ from $g_i$ $\forall i \in \{1\dots n\}\setminus\{\self\}$
	\STATE $m_{out} = \bigoplus_{i=1\dots n} M_i = \bigoplus_{i=1\dots n} \left(m_i \oplus \bigoplus_{j=1\dots n,j\not=i} s_{i,j} \right) = \bigoplus_{i=1\dots n} m_i $
\end{algorithmic}
\caption{Dining Cryptographer Protocol as executed by node $g_{\self}$.}
\label{alg:dcn1}
\end{algorithm}

Please note, that all secrets $s_{i,j}$ are symmetrical, i.e., \(s_{i,j} = s_{j,i}\), and are shared between pairs of nodes \(g_i,g_j.\)
The result \(m_{out} = \bigoplus_{i=1\dots n} M_i\) contains every index combination \(i\not=j\) exactly once.
Therefore, all secret pairs \(s_{i,j},s_{j,i}\) eliminate each other  \(s_{i,j}\oplus s_{j,i} = 0.\)

Dining-cryptographers protocols are a well-known privacy-preserving primitive for network communication.
They are applied in small groups of nodes in various modern protocols~\cite{dissent, dissent2, ahn, flex}.
Dissent~\cite{dissent,dissent2} applies them as its communication protocol in the core anonymity network.
Von Ahn et al.~\cite{ahn} and also we, in previous work~\cite{flex}, use them as group components to provide strong sender anonymity.
So their security properties are relevant for modern designs as well.

Using DC networks for implementing a broadcast will be very inefficient for large groups.
To mitigate this, a reasonably-sized sub-group could run a DC protocol. 
Some of the members then start a flood-and-prune broadcast to reach all other group members, e.g., as we laid out in~\cite{moedinger2020trustcom}.
However, care has to be taken on how the flood-and-prune phase is started so that it does not reveal the originator or the entire group composition.

\subsection{Shamir's Secret Sharing}

Lastly, we introduce Shamir's secret sharing~\cite{Shamir}.
The scheme splits a message into $n$ shares so that $k$ with $1 \leq k \leq n$ shares are required to reconstruct the original message.
This is often called a $(n, k)$ threshold scheme.

Any polynomial $f = \sum_{i=0}^{k-1} a_ix^i,a_{k-1}\not=0$ of degree $k-1$ is unambiguously defined by any $k$ points~\cite{Polynomial} and can be reconstructed from them.
Given $n$ distinct points of $f$ with $\forall i \ne j: x_i \ne x_j$, we can denote the set as:

\begin{equation}
\{(x_1, f(x_1)), (x_2, f(x_2)), \dots, (x_n, f(x_n))\}.
\end{equation}

The original polynomial can be recovered from any subset of points of size $k$.
Lagrange interpolation provides the formula to recover the original polynomial, which works over the real numbers as well as over fields $\mathbb{Z}_p$, making all operations over integers modulo $p$.
This leads to the same result independent of the chosen points~\cite{Polynomial} and is computed by:
\begin{align}
f(x) &= \sum_{i=1}^{k} f(x_i)\mathcal{L}_i(x), \\
\mathcal{L}_i(x) &= \prod_{j=1,j\not=i}^{k}\frac{x-x_j}{x_i-x_j}.
\end{align}

Given a message $m\in\mathbb{Z}_p$ we now want to construct a polynomial $f\in\mathbb{Z}_p[x]$, the polynomial space over the given integers.
The degree of $f$ is $\operatorname{deg}(f) = k-1$ and $f(0)=m.$
A polynomial can be constructed easily by choosing integers $r_1,\ldots,r_{k-1}\in\mathbb{Z}_p\backslash\{0\}$ randomly and computing

\begin{equation}
f(x) = m + \sum_{i=1}^{k-1} r_ix^i.
\end{equation}

It is easy to see that $f(0)=m$, as all other coefficients will be eliminated, and it holds that the degree of $f$ is $k-1$.
The required $n$ secret shares can then be computed as

\begin{equation}
s_i = (i, f(i)), i\in\{1,\ldots,n\}.
\end{equation}

A Galois field $\text{GF}(2^n)$ of suitable size is used to implement Shamir's secret sharing efficiently, usually $\text{GF}(2^8)$.
A notable property of these fields is that the addition of elements is equivalent to bitwise XOR of their binary representation.

\subsection{Goal}

Our honest peers' goal is to broadcast a message within the network while maintaining sender anonymity, i.e., at least $k-1$ other nodes should be indistinguishable from them as the originator, where $k$ depends on the parameters chosen in the system.
Honest nodes will strictly follow the protocol, as their goal is to broadcast messages correctly.

The primary goal of the attacker is to identify the participant sending the message.
Attackers follow the semi-honest model, i.e., they follow the protocol, with a small modification: They are allowed to refuse cooperation in the flood and prune broadcasting phase.
They will combine all knowledge they can acquire throughout the protocol, e.g., all messages they receive.
Attackers cannot manipulate the network, compromise other nodes, and solve computationally-infeasible problems such as encryption schemes.
The privacy section details additional measures and their applicability with malicious attackers.

\section{Secret-Shared Dining-Cryptographers Protocol}
\label{sec:scheme}

Within a large network, consider a group of size $n$, where one participant wants to transmit a message into the entire network.
We change the broadcast of the message to all participants into the transmission of $n$ distinct parts while still using a dining-cryptographers broadcast.
The parts are created using a $(n,k)$ Shamir's secret-sharing technique.
Each part is transmitted simultaneously during a modified dining cryptographer round, resulting in each participant ending up with a single share of the message.
The values of $k$ and $p$ required for the secret-sharing are system parameters, i.e., they are known beforehand and stay the same in the whole system.

\begin{figure}[ht]
	\centering
	\begin{tikzpicture}
	\node[state](split) {Split (Step~\ref{stsplit})};
	\node[state, below of=split, yshift=-0cm, fill={rgb:orange,1;yellow,2;pink,5}](bc) {Distribute (Steps~\ref{stbc1}, \ref{stbc2}, \ref{crstep}, \ref{steprec}, \ref{stbc3})};
	\node[state, below of=bc, yshift=-0cm](coop) {Broadcast and Combine (Step~\ref{stbc}, \ref{strc})};
	
	\draw [arrow, very thick] (split) -- (bc);
	\draw [arrow, very thick] (bc) -- (coop);
	\end{tikzpicture}
	\label{fig:steps}
	\caption{The three phases of the protocol and their corresponding steps explained in \Cref{alg:dcn2}. A message gets split up into $n$ shares, which are then distributed to group members via a dining-cryptographers broadcast, one share for each. Any $k$ members can then cooperate and recover the original message, corresponding to the split phase.}
\end{figure}
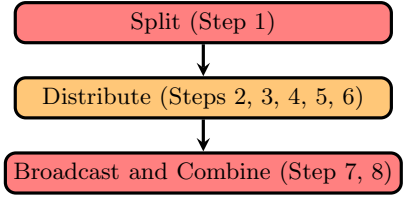

Our protocol consists of three phases, which are shown in~\Cref{fig:steps}.
In the first phase, named Split, a given message $m$ is split into $n$ secret shares.
To split the message, we chose $k-1$ random numbers to create a random polynomial $f$ which evaluates as $f(0) = m$.
Lastly we compute the secret shares $s_i = (i,f(i) \mod p)$ for all $i\in[1,n]$.

In the following phase (the distribution phase), each of the $n$ participants of the network then receives a unique share of the secret.
As described above, the DC protocol can only be used to make anonymous broadcasts, but not to send individual messages to certain participants anonymously. 
We can modify the protocol in such a way that this becomes possible.
The modified DC protocol version is shown in~\Cref{alg:dcn2}, note that a node that does not intend to send anything still proceeds with \(m_{\self}=0\).
Further note that paricipants not part of the group just execute Step~8 of~\Cref{alg:dcn2}.
The key modification compared to the original DC protocol as described by Chaum~\cite{dcns} (shown in \Cref{alg:dcn1}) is that Step~\ref{step4} no longer makes a broadcast but transmits individual messages to other participants.
The impact of this change is discussed in \Cref{sec:priv}.

\begin{algorithm}[ht]
\begin{algorithmic}[1]
	\renewcommand{\algorithmicrequire}{\textbf{Input:}}
	\renewcommand{\algorithmicensure}{\textbf{Output:}}
	\REQUIRE Participants $g_1, g_2, \dots, g_n$, message $m_{\self}$ of length $\ell$
	\ENSURE  Message $m_{\self,out}$, the message transmitted to this entity
	\STATE\label{stsplit} Split $m_{\self}$ into $n$ parts $m_{\self, 1}, \dots, m_{\self, n}$ using the secret-sharing scheme
	\STATE\label{stbc1} Establish shared random secrets $s_{\self,i}$ of length $\ell$ with each member $g_i,i\not=\self$
	\STATE\label{stbc2} $M_{\self,i} =m_{\self, i} \oplus \bigoplus_{j=1\dots n,j\not=\self} s_{\self,j}$ $\forall i \in \{1\dots n\}$
	\STATE\label{crstep} Send $M_{\self,i}$ to $g_i$ $\forall i \in \{1\dots n\}\setminus\{\self\}$
	\STATE\label{steprec} Receive $M_{i,\self}$ from $g_i$ $\forall i \in \{1\dots n\}\setminus\{\self\}$
	\STATE\label{stbc3} $m_{\self, out} =\bigoplus_{i=1\dots n,j\not=\self} M_{i,\self}$
	\STATE\label{stbc} Broadcast $m_{\self, out}$ to all network participants
	\STATE\label{strc} Reconstruct $m_{out}$ after receiving $k-1$ other shares
\end{algorithmic}
\caption{Modified DC protocol as executed by node $g_{\self}$.}
\label{alg:dcn2}
\end{algorithm}

The output of the distributed XOR function that participant $g_h$ computes is no longer $m_{out} = \bigoplus_{i=1\dots n} m_i $ but rather $m_{h, out} = \bigoplus_{i=1\dots n} m_{i,h}.$
Each member must now broadcast the message throughout the network.

If at least $k$ participants broadcast their message, every recipient can decode the original message.
If $k-2$ or fewer participants broadcast the message, no one can decode the message.
When exactly $k-1$ participants broadcast, only non-broadcasting participants of the group can decode the message, as they possess the last share required to decrypt the message themself.
Verifying the correctness of the result is omitted for the simplicity of the presentation.
It would require application-level integrity protection, i.e., there needs to be a way to ensure a message is valid for the application using the protocol.

\subsection{Correctness}

For the protocol's correctness, we assume all participants execute the DC protocol correctly, no errors occurred, and everyone used a $(n,k)$ Shamir's secret sharing technique.
In a first step, we show that participants can reconstruct the sum of all Shamir's secret sharing points from the messages received in the DC protocol.
From this, we reconstruct the original message $m_i\not=0$ in a second step, given a successful sharing round.

\subsubsection{Recovering the Sum of All Shared Points}

The $i$-th participant receives the $n-1$ messages $M_{1,i}\ldots M_{i-1,i}M_{i+1,i}\ldots M_{n,i}.$
Further, they create the message $M_{i,i}$ themselves. Each message has the form:

\begin{equation}
M_{h,i}=m_{h,i}\oplus \bigoplus_{j\in \{1\ldots n\}\setminus \{h\}} s_{h,j}.
\end{equation}

Therefore, the combination through XOR of all receives messages is

\begin{align}
\begin{split}
\bigoplus_{h\in \{1\ldots n\}} M_{h, i} &=\bigoplus_{h\in \{1\ldots n\}} \left(m_{h,i} \oplus \bigoplus_{j\in \{1\ldots n\}\setminus \{h\}} s_{h,j}\right)\\
 &=\left(\bigoplus_{h\in \{1\ldots n\}} m_{h,i}\right) \oplus \left(\underbrace{\bigoplus_{h\in \{1\ldots n\}} \bigoplus_{j\in \{1\ldots n\}\setminus \{h\}} s_{h,j}}_{=0, \text{ as }s_{h,j}\oplus s_{j,h} = 0}\right)\\
 &=\bigoplus_{h\in \{1\ldots n\}} m_{h,i}.
 \end{split}
\end{align}

As $m_{h,i}$ was created through the Shamir's secret sharing protocol, they have the form $m_{h,i} = p_h(i).$
Here $p_h$ is the polynomial created by participant $h$ to split their message.
The polynomial is created over the Galois field $\text{GF}(2^8),$ a field with characteristic 2.
In fields of characteristic 2, XOR and addition are equivalent.
Therefore, it holds that:

\begin{equation}
\bigoplus_{h\in \{1\ldots n\}} m_{h,i} = \bigoplus_{h\in \{1\ldots n\}} p_h(i) \stackrel{\text{over }\text{GF}(2^q)}{=} \sum_{h\in \{1\ldots n\}}  p_h(i).
\end{equation}

\subsubsection{Reconstruction of the Shared Message}

In this second step, we show that receiving $k$ distinct results allows us to reconstruct the protocol's original message input.
We assume that the flooding mechanism, or any appropriate sharing protocol, correctly distributed $k$ shares to all participants.
Without loss of generality, we assume a participant received the first $k$ messages:

\begin{equation}
\sum_{h\in \{1\ldots n\}} p_h(1), \ldots, \sum_{h\in \{1\ldots n\}} p_h(k).
\end{equation}

We saw in the section on Lagrange interpolation, that polynomial interpolation is uniquely possible with $k$ evaluation points $p(1),\ldots,p(k)$ for a polynomial $p$ of degree $deg(p)=k-1.$
We interpret our received messages as points of a polynomial $p_{\sum}:$

\begin{equation}
p_{\sum} (i) := \sum_{h\in \{1\ldots n\}} p_h(i).
\end{equation}

Polynomial interpolation is unique with the given degree restrictions, and polynomial addition cannot increase the degree of the resulting polynomial.
It holds, therefore, that:

\begin{equation}
p_{\sum} = \sum_{h\in \{1\ldots n\}} p_h.
\end{equation}

Evaluation and addition is commutative for polynomials, i.e., $(f+g)(x) = f(x) + g(x).$
Lastly, assume the messages are encoded at evaluation position s.

\begin{equation}
p_{\sum}(s) = \left(\sum_{h\in \{1\ldots n\}} p_h\right)(s) = \left(\sum_{h\in \{1\ldots n\}} \underbrace{p_h(s)}_{=m_i}\right)
\end{equation}

If at most one message $m_i\not=0$ exists, the reconstruction of the message is successful.
Otherwise, the sum of all non-zero messages is restored.

\section{Security and Privacy Evaluation}
\label{sec:priv}

We assume a group size of $n$ participants using a secure $(n,k)$-secret sharing scheme for this evaluation.
We restrict ourselves to group communication, as the flood and prune broadcast has no interesting privacy or security properties.

\subsection{Goal}

Let $M_i=(M_{i,1},\ldots,M_{i,n})$ be the vector of messages created by node $i$ in a system with $n$ participants.
Let $f$ be the function combining such a vector into the intended message, i.e., the combination algorithm of the secret sharing scheme.
Within the formalisation, we denote the previously presented \Cref{alg:dcn2} as $\operatorname{Alg\ref{alg:dcn2}},$ which is used to create all messages $M_{i,j}$.
Let the probability of $k-1$ attackers successfully identify a node sending a message be denoted by: 

\begin{equation}
P\left[f(M_{\ell})\not=0 \middle| 
\begin{matrix}
pp \leftarrow \operatorname{Setup}(\lambda, k, f) \\
M_i := M_{i,j}, i,j\in \{1\ldots n\} \leftarrow  \operatorname{Alg\ref{alg:dcn2}}(pp) \\
\ell\in\{1,k+1,\ldots,n\} \leftarrow A(pp,M_{i,j}, j\in \{2\ldots k\})
\end{matrix}\right].
\end{equation}

We call our scheme $(n,k-1)$ secure if this probability is only negligibly different from selecting a participant out of the $n-k+1$ non attackers at random, i.e., 

\begin{equation}
\left|P-\frac{1}{n-k+1}\right|<\operatorname{negl}(\lambda).
\end{equation}

Informally, this definition is true when $k-1$ colluding nodes cannot identify the originator of the message within the set of $n-|\text{attackers}|$ non-colluding nodes.
But once $k$ nodes cooperate, no guarantees are made.

\subsection{Semi-Honest Model}

To show our scheme fulfills the previous definition, let there be $k-1$ colluding attackers present in the group, which follow the semi-honest model.
Assume, without loss of generality as the nodes can be renumbered, that the victim has index $1$ and the attackers' index 2 through $k.$

These colluding participants can collect $k-1$ messages $M_{i,j}$ of the form $M_{i,j} = m_{i,j} \oplus \bigoplus_{h\in\{1\ldots n\}} s_{i,h}$ by any participant $i$ and the honest reconstruction of $ p_{\sum},$ which provides the transmitted message $m$ and the sum of all point evaluations.
To identify the originator, the attackers need to compute any $m_{1,j}$ of the victim or, equivalently, their aggregate key $\bigoplus_j s_{1,j}.$
The original proof of Chaum holds for directly reconstructing $\bigoplus_j s_{1,j},$ so we will focus on $m_{1,j}.$
Note that $m_{1,j}=p_1(j)$ is equivalent, where the polynomial $p_1$ has degree $deg(p_i)=k-1$ and the form

\begin{equation}
p_i(x) = \sum_{\ell=1}^{k} a_\ell x^{\ell-1}.
\end{equation}

Given $k-1$ messages $M_{1,2}\ldots M_{1,k+1}$ and $i\not=j$ we can see that it holds that

\begin{align}
\begin{split}
M_{1,i} \oplus M_{1,j} 
&= \left(m_{1,i} \oplus \bigoplus_{h\in\{1\ldots n\}} s_{1,h}\right) \oplus \left(m_{1,j} \oplus \bigoplus_{h\in\{1\ldots n\}} s_{1,h}\right) \\
&= m_{1,i} \oplus m_{1,j} \oplus \left(\bigoplus_{h\in\{1\ldots n\}} s_{1,h} \oplus \bigoplus_{h\in\{1\ldots n\}} s_{1,h}\right) \\
&= m_{1,i} \oplus m_{1,j} \oplus \bigoplus_{h\in\{1\ldots n\}} \left(\underbrace{s_{1,h} \oplus s_{1,h}}_{=0}\right)\\
&= m_{1,i} \oplus m_{1,j}
\end{split}
\end{align}

As XOR and addition are equivalent over base fields of characteristic 2, which we use, and that $m_{i,j} = p_i(j),$ we can see that

\begin{equation}
m_{1,i} \oplus m_{1,j} = p_1(i) + p_1(j).
\end{equation}

Note that this only holds for even combinations, i.e., we cannot create $p_1(2) + p_1(3) + p_1(4).$
All combinations with an even number of parts can be constructed as a linear combination of combinations of two parts.
Therefore, using this equation, we can create only $k-2$ linearly independent equations:

\begin{equation}
[p_1] = 
\begin{cases}
\begin{matrix}
	\sum_{i=1}^{k} 2a_i2^{i-1}3^{i-1} &=& p_1(2) + p_1(3) \\
	\vdots && \vdots \\
	\sum_{i=1}^{k} 2a_i(k-1)^{i-1}k^{i-1} &=& p_1(k-1) + p_1(k) \\
\end{matrix}
\end{cases}
\end{equation}

The attackers can reconstruct the transmitted message \(m=p_{\sum}(0)\) by following the protocol normally.
Removing all attacker polynomials $p_2 \ldots p_k$ leaves

\begin{equation}
p_{\sum} - \sum_{j=2}^k p_j = p_1 + \sum_{j=k+1}^n p_j =: p_{\text{remains}}.
\end{equation}

Using this and applying the strategy to compute \([p_1]\) on all non-colluding participants allows the attackers to create the following matrix

\begin{equation} 
\begin{bmatrix}
	[p_1] & [0] & \cdots & [0] & S_1 \\
	[0] & [p_{k+1}] &  & [0] & S_{k+1}\\
	\vdots &  & \ddots{} & & \vdots \\
	[0] & [0] &  & [p_{n}] & S_{n}\\
	1\ldots1 & 1\ldots1 & \ldots & 1\ldots1 & p_{\text{remains}} \\
\end{bmatrix}
\end{equation}

All entries $[p_i]$ represent the previous equation systems with their respective solution vectors $S_i = (p_i(2)+p_i(3),\ldots,p_i(k-1)+p_i(k))$ generated from the messages $M_{i,j}.$
Each block $[p_i]$ and $[0]$ have $k-2$ rows, while the final row models $p_{\text{remains}},$ where all coefficients are present exactly once.
All further derivations of $p_{\text{remains}}$ would not be linearly independent equations.
There is no further relation between the remaining polynomials $p_1,p_{k+1}, \ldots, p_n,$ as all are chosen independently at random.

Solving the equations for a single participant leaves us with \(k-2+1\) rows ($[p_i]$ and $p_{\text{remains}}$) and $k$ indeterminants $a_1,\ldots,a_k$ and therefore $k$ columns.
The full matrix has \((n-k+1)\times(k-2)+1\) rows and \(k\times (n-k+1)+1\) columns.
Using the Rouché–Capelli theorem, i.e., if for a system of equations $Ax=b$ there is a unique solution iff $rank(A)=rank(A|b)$, this results in infinitely many solutions, i.e., ambiguous reconstruction, and further breaks the security assumption of the base secret sharing protocol.

If a message can be verified after decryption, an exhaustive search for solutions is possible.
The underlying field size determines the cost for an exhaustive search, i.e., the field size corresponds to $\lambda$ in our previous definition.
Absent any notes identifying correct solutions, all solutions to the system of equations are equally valid and likely, i.e., any of the \(n-k+1\) possible victims might be the sender with equal probability \(P[f(M_\ell)\not=0]=\frac{1}{n-k+1}.\)

\subsection{Outside Observers}

Outside observers cannot determine the origin of a broadcast as long as secure channels are used, as all participants have to send data of the same size for each transmission.
Similar to classical DC networks, no guarantees can be retained when the channels are no longer secure.

\subsection{Modern DC Malicious Mitigations}

While attackers act semi-honest in the previous evaluation, modern dining-cryptographers protocols apply various mechanisms to deal with collisions, fairness, and robustness issues of the protocol~\cite{golle, ahn}.
$2n$ slots are used to increase fairness, where every participant may use at most one slot at a time, which they chose randomly.
For each secret share, a commitment is created and broadcasted to the group.
A zero-knowledge proof is used when more than half of all slots are used.
The proof shows that a participant used at most one slot.
Lastly, the most problematic case, selective non-participation, can be combated by pre-emptively sharing all secrets in encrypted form with the group.

These techniques can be applied to our proposed protocol to make it resistant to malicious participants.
Slots can be easily introduced by applying the secret splitting per slot, not on the full message.
Commitments can be created in the same form as by von Ahn et al.~\cite{ahn}: each slot provides its own commitments.
The zero-knowledge proof of fairness by von Ahn et al. can be extended as easily: The opening of commitments is combined with a reconstruction of the secret shares into the actual message.
This message has to be zero.

\section{Performance Evaluation}
\label{sec:perf}

This section shows the performance results for our scheme and the methodology used to acquire those results.

\subsection{Methodology}

We implemented a prototype simulation that can simulate both the original DC protocol and our modified version. 
The simulation is available online\footnote{See \url{https://github.com/vs-uulm/thc-in-dc-simulation}} and written in Java.
We use built-in synchronization utilities to model the communication and synchronization of participants.
For threshold cryptography, we used the open-source library shamir\footnote{\url{https://github.com/codahale/shamir}} in version 0.7.0.
The Shamir library uses a Galois field $\text{GF}(2^8)$ as a base field.
The library provides two methods, split and join, of combined complexity of $\mathcal{O}( \ell\cdot (n+k^2)) $.

We ran this implementation 10 to 30 times for each combination of parameters.
We aggregated the measured throughput and computed the average and standard deviation.

Network latency is simulated, but we set it to $0$ to prevent influence on the measured variable when not specified.
To mitigate our results' distortion due to runtime optimization attempts by the Java virtual machine, we ran a warm-up phase before each test.
In this warm-up phase, $100$ runs were performed that are not included in our results.

We compared the modified DC protocol, denoted as Broadcast, to Chaum's original version's performance, denoted as DC Phase in graphs.
We investigated several core issues: 
\begin{itemize}
	\item The size $\ell$ of the transmitted message,
	\item the scaling behaviour of the protocol, i.e., increasing $n$,
	\item the performance impact of variable $k$ values,
	\item the influence of network latency.
\end{itemize}

For the performance evaluation, we consider a simple collaboration protocol in place of the broadcast to reduce simulation effort.
Participants collaborate with at least $k-1$ other members to recover the original message $m$.
\Cref{alg:coop} provides a cooperation scheme which, when executed correctly by each network member, produces a minimal amount of transmitted messages: $n(k-1).$ 

\begin{algorithm}[ht]
\begin{algorithmic}[1]
	\renewcommand{\algorithmicrequire}{\textbf{Input:}}
	\renewcommand{\algorithmicensure}{\textbf{Output:}}
	\REQUIRE Message part $m_i:=\bigoplus_j m_{j,i}$, Group members $g_1, g_2, \dots, g_n$, number of required message shares $k$\\
	\ENSURE Message $m_{out}$ \\
	\STATE Send $m_i$ to $g_{j}\  \forall j \in \{i+a \mod (n+1) \mid a\in \mathbb{N}, 1 \leq a \leq k-1 \}$
	\STATE Receive $m_r$ from $g_r \forall r \in \{ x \mid \exists a, 1 \leq a \leq k-1 : x+a \mod (n+1) = i\}$ 
	\RETURN $m_{out}$ from the $k-1$ received messages and $m_i$.
\end{algorithmic}
	\caption{Combine protocol to emulate broadcast.}
\label{alg:coop}
\end{algorithm}

We opted not to evaluate a full flooding approach, as this would shift the focus from the modifications we performed.
Additionally, the performance characteristics of flooding approaches are well known.

\subsection{Message Size \texorpdfstring{$\ell$}{l}}
\label{sec:msgsize}

Both the original DC protocol and our modified protocol transmit a message of the fixed-length $\ell$ each round.
We want to keep $\ell$ as close as possible to the actual length of the information we want to send.

Messages longer than $\ell$ can be split into multiple messages, increasing overhead and, therefore, decreasing throughput.
If the information is shorter than $\ell$, it can be padded with $0$-bytes to make it size $\ell$, leading to the transmission of more data than necessary, producing overhead as well.

\begin{figure}[ht]
\centering
	\begin{tikzpicture}
	\begin{axis}[
	width=.5\textwidth,
	/pgf/number format/.cd,
	1000 sep={},
	xlabel={Bytes transmitted per round ($\ell$)},
	ylabel={Throughput [kB/s]},
	ymin=0, ymax=1000,
	legend pos=north west,
	ymajorgrids=true,
	grid style=dashed,
	xmode=log,
]

	\addplot[
	color=blue,
	mark=o,
	only marks,
	]
	plot [error bars/.cd, y dir = both, y explicit]
	table [col sep=comma, y error index=2] {data/final_bm/IncreasingBPR_D10.csv};
	\addplot[
	color=red,
	mark=square,
	only marks,
	]
	plot [error bars/.cd, y dir = both, y explicit]
	table [col sep=comma, y error index=2] {data/final_bm/IncreasingBPR_S10.csv};
	\legend{DC Phase, Broadcast~}

	\end{axis}
	\end{tikzpicture}
	\caption{The measured throughput and its standard deviation while increasing $\ell$ for $n=10$ and $k=3$. The transmitted information has size \SI{8}{\kilo\byte}. Performance reaches its peak when $\ell$ is about the size of the transmitted information.}
	\label{fig:incbpr2}
\end{figure}
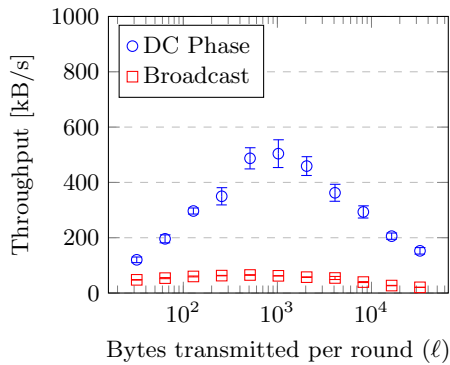

We show the results of this overhead in~\Cref{fig:incbpr2}.
We varied $\ell$ from $\SI{32}{\byte}$ to $\SI{32}{\kilo\byte}$ with $n=10$ and a given real message size of $\SI{8}{\kilo\byte}$.
We chose the relevant parameters for this benchmark with regard to the potential use for our proposed system in the field of cryptocurrencies.
Therefore we picked sizes applicable to groups~\cite{flex} and transaction sizes, validating our assumption that the performance is at its peak when $\ell$ is roughly equal to the size of the information to transmit.
Results for varying sizes of $n$ (not shown) lead to the same validation.

\subsection{Network Size \texorpdfstring{$n$}{n}}

While the message complexity for the core DC protocol is identical in both schemes, a round of the modified version of the DC protocol needs additional messages in the cooperation phase.
When keeping $k$ constant, the modified version of the protocol requires $\mathcal{O}(n)$ more transmissions than the original protocol.
As we see in \Cref{fig:incn3}, this makes a significant difference for a low number of participants.
Because both versions of the protocol are of overall complexity $\mathcal{O}(n^2)$, the linear performance penalty becomes less of a concern when $n$ grows larger.

The increased number of sent messages is only one reason for the worse performance of our system.
The time for performing one round of the protocol can be divided into two parts:
time spent communicating and time spent for calculations.
Our system requires a larger amount of computations compared to the classical DC protocol.
In addition to performing the core DC functionality, it also splits and joins the messages to transmit using a secret-sharing scheme, resulting in the strong performance difference seen in \Cref{fig:incn3}.

\subsection{Network Delay}

\begin{figure}[htbp]
\centering
	\subfloat[Delay $0$ms]{
		\begin{tikzpicture}
		\begin{axis}[
		width=.5\textwidth,
		/pgf/number format/.cd,
		1000 sep={},
		xlabel={Number of participants},
		ylabel={Throughput [kB/s]},
		legend pos=north east,
		ymajorgrids=true,
		grid style=dashed,
		ymode=log,
		]
		
		\addplot[
		color=blue,
		mark=o,
		only marks,
		]
		plot [error bars/.cd, y dir = both, y explicit]
		table [col sep=comma, y error index=2] {data/final_bm/IncreasingN_D.csv};
		\addplot[
		color=red,
		mark=square,
		only marks,
		]
		plot [error bars/.cd, y dir = both, y explicit]
		table [col sep=comma, y error index=2] {data/final_bm/IncreasingN_S.csv};
		\legend{DC Phase, Broadcast~}
		
		\end{axis}
		\end{tikzpicture}
		
		\label{fig:incn3}
	}
	\subfloat[Delay $100$ms]{
		\begin{tikzpicture}
		\begin{axis}[
		width=.5\textwidth,
		/pgf/number format/.cd,
		1000 sep={},
		xlabel={Number of participants},
		ylabel={Throughput [kB/s]},
		legend pos=south west,
		ymajorgrids=true,
		grid style=dashed,
		ymode=log,
		]
		
		\addplot[
		color=blue,
		mark=o,
		only marks,
		]
		plot [error bars/.cd, y dir = both, y explicit]
		table [col sep=comma, y error index=2] {data/final_bm/IncreasingN_DDDD.csv};
		\addplot[
		color=red,
		mark=square,
		only marks,
		]
		plot [error bars/.cd, y dir = both, y explicit]
		table [col sep=comma, y error index=2] {data/final_bm/IncreasingN_SDDD.csv};
		\legend{DC Phase, Broadcast~}
		
		\end{axis}
		\end{tikzpicture}
		\label{fig:incn4}
	}
	\caption{Measuring throughput in DC protocol runs over networks of various sizes. Variable $n$, $k=3$, $\ell=$\SI{8}{\kilo\byte}.}
\end{figure}
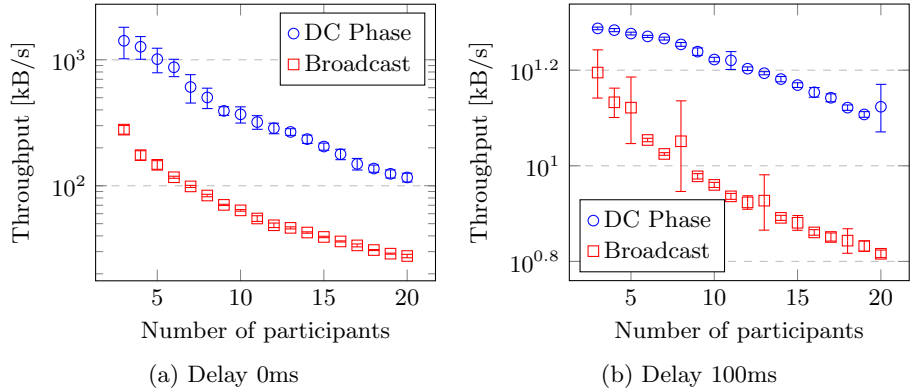

To investigate less optimal environments, we simulated our scheme using a delay of $>0$ms.
The gap in performance between the original DC protocol and our system is notably smaller.
The results of adding a delay of \SI{100}{\milli\second} are shown in \Cref{fig:incn4} respectively, but simulations with intermediate values show similar results.
We chose \SI{100}{\milli\second} as a typical representation of internet communication delay, but in real-world scenarios, it can be considerably smaller.

Note that when adding delay, our system only improves relative to the original dining-cryptographers protocol. 
The absolute performance of both approaches suffers under message transmission delay.
We measured throughput rates of \SI{13.58}{\kilo\byte/\second} for $n=4$ and \SI{9.12}{\kilo\byte/\second} for $n=10$ with a delay of \SI{100}{\milli\second}.

\subsection{Number of Shares \texorpdfstring{$k$}{k}}

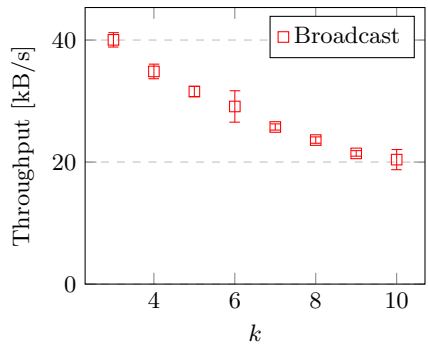
\begin{figure}[htbp]
\centering
	\begin{tikzpicture}
	\begin{axis}[
	width=.5\textwidth,
	/pgf/number format/.cd,
	1000 sep={},
	xlabel={$k$},
	ylabel={Throughput [kB/s]},
	ymin=0,
	legend pos=north east,
	ymajorgrids=true,
	grid style=dashed,
	]

	\addplot[
	color=red,
	mark=square,
	only marks,
	]
	plot [error bars/.cd, y dir = both, y explicit]
	table [col sep=comma, y error index=2]{data/final_bm/IncreasingK_S10.csv};
	\legend{Broadcast}

	\end{axis}
	\end{tikzpicture}
	\caption{Measuring network throughput while varying parameter $k$. The other parameters are kept constant with $n=10$ and $\ell$ as well as the size of the transmitted information as \SI{8}{\kilo\byte}.}
	\label{fig:inck}
\end{figure}

Lastly, the value of $k$ is the number of message shares needed to restore a message and significantly impacts the protocol.
This impact is due to participants needing to compute additional methods and perform additional $k-1$ transmissions to receive the shares.
\Cref{fig:inck} shows the results of benchmarking our system with $n=10$, $\ell=\SI{8}{\kilo\byte}$ and $k \in \{3,\dots,10\}$.
As expected, increasing $k$ decreases our system's performance.

\section{Applications}
\label{sec:app}

As we have seen, our version of a DC protocol typically achieves throughput rates between $\SI{10}{\kilo\byte/\second}$ and $\SI{100}{\kilo\byte/\second}$.
A real-world application for our system lies in the anonymous transmission of transaction data for blockchains, e.g., in an environment like the one proposed in~\cite{flex}.
Such transaction data are typically of size~$<\SI{1}{\kilo\byte}$, whereas group sizes are between $n=4$ and $n=10$ and transmission delay is around $\SI{100}{\milli\second}$.

Many blockchain systems produce only a few transactions per second, despite thousands of nodes participating in the network.
Separating these into groups for privacy is unlikely to lead to any groups that require more than one transaction per second.
Therefore, every system that can achieve speeds of~$>\SI{1}{}$ transactions made per second is suitable for application in a system as the one proposed in~\cite{flex}.
Our system is well-suited for such a task, as it can efficiently work with this load.

\section{Conclusion}
\label{sec:cnc}

In this work, we proposed a combination of the classical dining-cryptographers protocol and Shamir's secret sharing to enforce anonymity during a broadcast process.
This problem arises during a broadcast, as nodes that already received the message might refuse further cooperation.
We showed that the protocol is computational secure in the number of shares $k$, maintaining $n-|attackers|$-anonymity for at most $k-1$ attackers.

Our system provides a first, unoptimized solution, so further work could improve the system's performance and flexibility.
In our simulation, this results in throughput rates between $\SI{10}{\kilo\byte/\second}$ and $\SI{100}{\kilo\byte/\second}$ for a full broadcast simulation and over $\SI{500}{\kilo\byte/\second}$ with reasonable privacy settings.
These performance results show our system is viable for a wide range of applications, such as blockchain-transaction dissemination in peer-to-peer networks and other peer-to-peer applications requiring high privacy guarantees.

\bibliographystyle{splncs04}
\bibliography{bibliography}

\end{document}